\documentclass[pre,reprint,amssymb,amsmath,amsfonts,aps,superscriptaddress]{revtex4-1}
\usepackage{amsmath}
\usepackage{mathrsfs}
\usepackage[colorlinks=true,linkcolor=blue,citecolor=blue,urlcolor=blue]{hyperref}
\usepackage[suffix=]{epstopdf}
\usepackage{graphicx}
\begin{document}
\title{Dynamics of Rotator Chain with Dissipative Boundary}
\author{Pu Ke}
\author{Zhigang Zheng}
\email{zgzheng@bnu.edu.cn}
\affiliation{Department of Physics and the Beijing -- Hong Kong -- Singapore
  Joint Center for Nonlinear and Complex Systems (Beijing), Beijing
  Normal University, Beijing 100875, People's Republic of China}
\begin{abstract}
  We study the deterministic dynamics of rotator chain with purely
  mechanical driving on the boundary by stability analysis and
  numerical simulation.  Globally synchronous rotation, clustered
  synchronous rotation, and split synchronous rotation states are
  identified.  In particular, we find that the single-peaked variance
  distribution of angular momenta is the consequence of the
  deterministic dynamics.  As a result, the operational definition of
  temperature used in the previous studies on rotator chain should be
  revisited.
\end{abstract}
\maketitle
\section{INTRODUCTION}
\label{sec:intro}
The violation of Fourier's law in low dimensional lattice has drawn
much attention in recent years due to its fundamental importance to
non-equilibrium thermodynamics and statistical
mechanics\cite{Dhar2008}.
This phenomenological law, which relies on the local equilibrium
hypothesis, is a macroscopic description of non-equilibrium process,
and has been verified to be accurate through various experimental
settings.  However, a rigorous derivation of Fourier's Law from
microscopic statistical-mechanical argument is still missing, which
motivated a large number of studies on energy conduction in various
models.  As an unexpected result drawn from these studies, Fourier's
Law is violated for the divergence of heat conductivity with system
size in many one-dimensional oscillator-based systems, unless the
substrate potential
exists\cite{PhysRevLett.78.1896,PhysRevLett.52.1861,PhysRevE.67.041205}
or the interaction potential is asymmetrical\cite{PhysRevE.85.060102}.

The model of rotator chain was introduced as a counter example to the
oscillator-based models.  Even without any substrate potential, a 1D
rotator chain model, whose interaction potential is also symmetric
exhibits normal heat
conduction\cite{PhysRevLett.84.2381,PhysRevLett.84.2144}.
Nevertheless, when both thermal and mechanical driving exist in this
model, the variance profile of momenta, which is commonly used as the
operational definition for temperature in the system that involves
translational motion\cite{PhysRevE.84.061108}, is
single-peaked\cite{PhysRevE.84.061108}, this finding seems to directly
contradicts Fourier's Law since a necessary condition for a 1D system
to comply the law is to have a linear temperature distribution.  The
origin of this single-peaked distribution was thought to be the
interaction between the mechanical forcing and the thermal
forcing\cite{PhysRevE.84.061108}.  Nevertheless, little about the
variance of momenta is known when mechanical driving exists
exclusively.  Although rotator chain model is a special case of the
classical XY model, which drawn much attention in the fields of both
extensive\cite{PhysRevB.16.4945} and
non-extensive\cite{PhysRevLett.80.5313} statistical mechanics, few
studies have been focused on the deterministic dynamics of rotator
chain, especially the variance profile of angular momenta.  Given the
interaction potential of rotator chain differs from the
oscillator-based models by its periodicity and boundedness,
investigation on deterministic dynamics of rotator chain is necessary
to further develop the understanding of the energy conduction
properties of rotator chain.

The present paper focuses on the deterministic dynamics of the rotator
chain with purely mechanical driving.  We proved the
existence of the globally synchronous rotation and identified two
other dynamical states as the parameters vary.  The momenta variance
profile was observed to be qualitatively similar to the case when both
thermal and mechanical driving are presented.  As further
investigation, we also analyzed the momenta distribution, detailed
evolution information, and phase portrait of the interface rotators.
The results provided strong evidences that the nonlinear variance
profile of angular momenta was of deterministic origin.

The paper is organized as follows.  In Sec.~\ref{sec:model}, a
detailed description of rotator chain is presented and the existence
of the globally synchronous state is proved.  Sec.~\ref{sec:profiles} discusses
energy current, averaged momenta and momenta variance profiles.
Sec.~\ref{sec:states} is devoted to an explanation to the
nonlinearity of variance profiles by investigating rotation states of
interface rotators.  Finally, Sec.~\ref{sec:conclusions} concludes
the paper.
\section{ROTATOR CHAIN}\label{sec:model}

\subsection{Description of the Model}
\label{sec:description_model}
\begin{figure}[hbt!]
\includegraphics[trim = 350px 100px 310px
80px,clip,width=.5\textwidth]{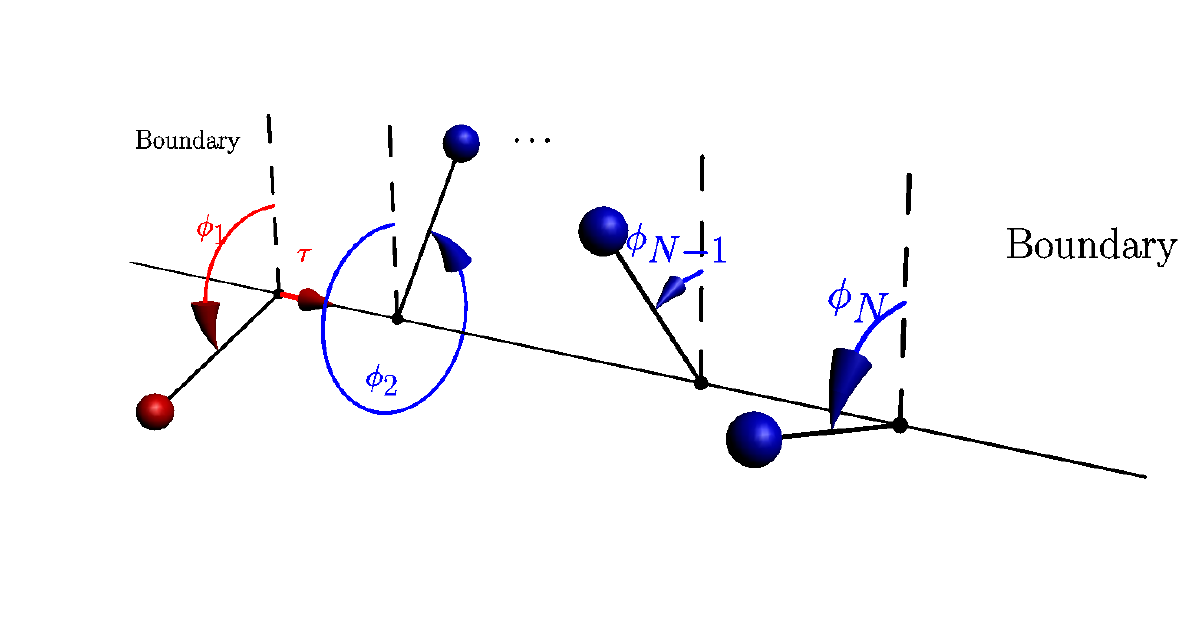}
  \caption{(Color online) Illustration of a rotator chain.}\label{fig:model} 
\end{figure}

A chain of $N$ rotators is described by the angles
$\boldsymbol{\phi}=(\phi_1,\phi_2,\dots,\phi_N)$ and their conjugate
angular momenta $\boldsymbol{L}=(L_1,L_2,\dots,L_N)$, See
Fig~\ref{fig:model}.
The nearest-neighbor interaction potential is periodical and takes the
form
\begin{equation}
  \label{eq:potential}
  U(\phi_{k},\phi_{k+1})=\epsilon\left[1-\cos(\phi_{k+1}-\phi_k)\right],
\end{equation}
where $\epsilon$ is the coupling coefficient.

Moreover, the system is homogeneous, that is, the moments of inertia of
all rotator are $I$.  Therefore, the Hamiltonian of the system is
\begin{equation}
  \label{eq:hamiltonian}
  \mathscr{H}(\boldsymbol{\phi},\boldsymbol{L})=\sum_k\left\{\frac{L_k^2}{2I}+\epsilon\big[1-\cos(\phi_{k+1}-\phi_{k})\big]\right\}.
\end{equation}
Dissipation is introduced on the both end for consistency to the case when
thermal driving exists (fluctuation-dissipation theorem).  Introducing
dimensionless time $s=\sqrt{\epsilon/I} t$ and letting
$\omega_k=\frac{\mathrm d\phi_k}{\mathrm ds}$, the dimensionless
equations of motion with open boundary reads
\begin{align}
  \notag\frac{\mathrm d\phi_k}{\mathrm ds}&=\omega_k,\\
  \notag\frac{\mathrm d\omega_k}{\mathrm
    ds}&=\sin(\phi_{k+1}-\phi_{k})-\sin(\phi_{k}-\phi_{k-1}),&& k\neq
  1,N\\
  \notag\frac{\mathrm d\omega_1}{\mathrm
    ds}&=\sin(\phi_{2}-\phi_{1})-\alpha\omega_1+\tau,\\
  \frac{\mathrm d\omega_N}{\mathrm ds}&=-\sin(\phi_N-\phi_{N-1})-
  \alpha\omega_N,\label{eq:evo_simplified}
\end{align}
where $\alpha=\gamma\sqrt{I/\epsilon}$ is the effective dissipation
coefficient and $\tau=\rm T/\epsilon$ is the effective torque, in which $\gamma$
is the dissipation coefficient and $\rm T$ is the mechanical torque.

\subsection{Dynamical States of Rotator Chain}
\label{sec:dyna_state}

Based on our analyses and observations, depending on
$\alpha$ and $\tau$, the rotator chain has three kinds of dynamical
states: \emph{globally synchronous rotation state}, \emph{split
  synchronous rotation state} and \emph{clustered synchronous state}.

In the globally synchronous rotation states, all rotators rotate at
the same angular momentum.  It can be proved that (see
Sec.\ref{sec:boundary}) for any $\alpha$ and $\tau<2$, i.e. the
driving torque is weak while the coupling is strong, the globally
synchronous rotation state is unique, and the phase difference between
every adjacent rotator pair is $\pi-\mathrm{arcsin}(-\tau/2)$
(cf. \eqref{eq:solution_phase_difference}).

In the split synchronous rotation state, the driving rotator rotates as if the
remaining rotators do not exits, while the remaining rotators rotate
slowly in synchronization.  This state could be attained when driving torque
is sufficiently large.

In the clustered synchronous rotation state, the entire chain is
divided into three region: fast rotation region, slow rotation region
and interface region.  In the fast and slow rotation region, the phase
difference between the adjacent rotators oscillates around $0$ and could not
exceed $2\pi$, while the rotators in the interface region exhibit
complicated dynamical pattern --- their rotations are affected by the
rotators both in fast rotation region and slow rotation region.

As a preliminary investigation, it is instructive to consider the
stability of the system analytically.  In the next
subsection, the uniqueness of stable fixed point of a rotator chain
without dissipation is proved and a saddle-node bifurcation is
identified; then the argument is extended to a rotator chain with
dissipative boundary to reveal the necessary condition $\tau>2$
(driving torque is large while coupling is weak) for
the forming of nonlinear momenta variance profile.

\subsubsection{Rotator Chain without Dissipative Boundary}
\label{sec:interior}
Having introduced phase difference $\delta_k=\phi_{k+1}-\phi_{k}$ and its
derivative $\Delta_k=\frac{\mathrm d\delta_k}{\mathrm ds}$, the equations
of motion can be cast into ($\alpha=0$)
\begin{align}\label{eq:diff}
  \notag \frac{\mathrm d\delta_k}{\mathrm ds}&=\Delta_k,\\
  \notag \frac{\mathrm d\Delta_k}{\mathrm
    ds}&=-2\sin\delta_k+\sin\delta_{k-1}+\sin\delta_{k+1},&&k\neq
  1,N-1\\
  \notag \frac{\mathrm d\Delta_1}{\mathrm
    ds}&=-2\sin\delta_1+\sin\delta_2-\tau,\\
  \frac{\mathrm d\Delta_{N-1}}{\mathrm
    ds}&=-2\sin\delta_{N-1}+\sin\delta_{N-2}.
\end{align}
The general formula of $\frac{\mathrm d\Delta_k}{\mathrm ds}=0$ is
$\sin\delta_k=k\sin\delta_1$, then the only possible solution
is $\sin{\delta_k}=0$ (i.e. $\delta_k=n_k\pi$). The Jacobian
matrix of \eqref{eq:diff} is
    
\begin{equation}
  \label{eq:jacobian_general}
  \mathbf{J}=\begin{pmatrix}
    \mathbf{0}&\mathbf{I}_{N-1}\\
    \mathbf{C}&\mathbf{0}
  \end{pmatrix},
\end{equation}
where $\mathbf{I}_{N-1}$ is a $N-1^{\mathrm{th}}$ order unit matrix, and $\mathbf{C}$ is
\begin{equation}
  \label{eq:jacobian_essential}
\begin{pmatrix}
  -2\chi_1&\chi_2&0&0&\cdots&0\\
  \chi_1&-2\chi_2&\chi_3&0&\cdots&0\\
  0&\chi_2&-2\chi_3&\chi_4&\cdots&0\\
  \vdots&\vdots&\vdots&\ddots&\ddots&\vdots\\
  0&0&0&0&\chi_{N-2}&-2\chi_{N-1}
\end{pmatrix},
\end{equation}
where $\chi_k=\pm 1$. The eigenequation of Jacobian Matrix is
\begin{align}
  \label{eq:eigenfun_J}
  \notag &\mathrm{det}
  \begin{pmatrix}
    -\lambda\mathbf{I_{N-1}}&\mathbf{I}_{N-1}\\
    \mathbf{C}&-\lambda\mathbf{I_{N-1}}
  \end{pmatrix}
  \\\notag &=\mathrm{det}(-\lambda
  \mathbf{I}_{N-1})
  \times\mathrm{det}(-\lambda\mathbf{I}_{N-1}+\mathbf{C}\frac{1}{\lambda})\\
 &=(-1)^{N-1}\mathrm{det}(\mathbf{C-\lambda^2\mathbf{I}_{N-1}})=0.
\end{align}
Which implies the eigenvalues of $\mathbf{J}$ are square roots of the
eigenvalues of $\mathbf{C}$.

If all $\chi_k=1 (\delta_k=2n_k\pi)$, then the eigenvalues of
$\mathbf{C}$ are $\lambda^2=-2(1+\cos\frac{k\pi}{N})<0$ $(
k=1,\cdots,N-1)$\cite{Kulkarni199963}. Since $\lambda$s are purely imaginary,
$\delta_k=2n_k\pi$ corresponds to a center in phase space.

Suppose that at least one $\chi_r=-1 (r<N-1,\delta_r=(2n+1)\pi)$, then
from Gershgorin's Theorem\cite{HornJohnson201210}, for eigenvalue
$\lambda_r^2$ of $\mathbf{C}$, $|\lambda_r^2-2|\leq
2$.  Because matrix $\mathbf{C}$ is nonsingular (all columns
are linear independent), then $\mathrm{Re}\lambda_r>0$, which renders the fixed
point unstable.  The argument still holds when more than one
$\chi=-1$, since the sum of all off-diagonal elements on arbitrary row
is smaller than $2$.

As proved above, the only stable fixed point of equation
\eqref{eq:diff} is $\delta_k=2n_k\pi$, which is a center.  The
corresponding motion is the collective rotation with the phase difference and
its derivative of each pair of rotator oscillate around
$(2n_k\pi,0)$ (libration).

If driving torque is added at the boundary, the general formula
of $\frac{\mathrm d\Delta_k}{\mathrm ds}=0$ becomes
$\sin\delta_k=k\sin\delta_1+(k-1)\tau$.  Substituting it into
$\frac{\mathrm d\Delta_{N-1}}{\mathrm ds}=0$ gives the necessary
condition for the existence of fixed points
$\sin\delta_1=(1-N)/N\tau$.  At the thermodynamic limit, $\displaystyle\lim_{N\to\infty}\sin\delta_1=-\tau$, and
$\tau=1$ is a saddle-node bifurcation point. 
\subsubsection{Rotator Chain with Dissipative Boundary}
\label{sec:boundary}
The role of the boundary rotators is of crucial importance in the
formation of nonlinear variance profile.  Consider a system consists
of only two boundary rotators
\begin{align}
  \label{eq:diff_boundary}
  \notag\frac{\mathrm d\delta}{\mathrm ds}&=\Delta,\\
  \frac{\mathrm d\Delta}{\mathrm
    ds}&=-2\sin\delta-\alpha\Delta-\tau.
\end{align}
Introducing new time scale $\xi=\sqrt{2}s$, then the equation
transform to
\begin{equation}
  \label{eq:josephson}
  \frac{\mathrm d^2\delta}{\mathrm
    d\xi^2}=-\sin\delta-\eta\frac{\mathrm
    d\delta}{\mathrm d\xi}-\beta,
\end{equation}
where $\eta=\frac{\sqrt{2}}{2}\alpha$ and $\beta=\frac{\tau}{2}$.  Despite the minus sign, equation
\eqref{eq:josephson} is the governing equation for the Josephson
junction, and received extensive study\cite{levi1978dynamics}.
\begin{figure}[hbt!]
\includegraphics[width=.3\textwidth]{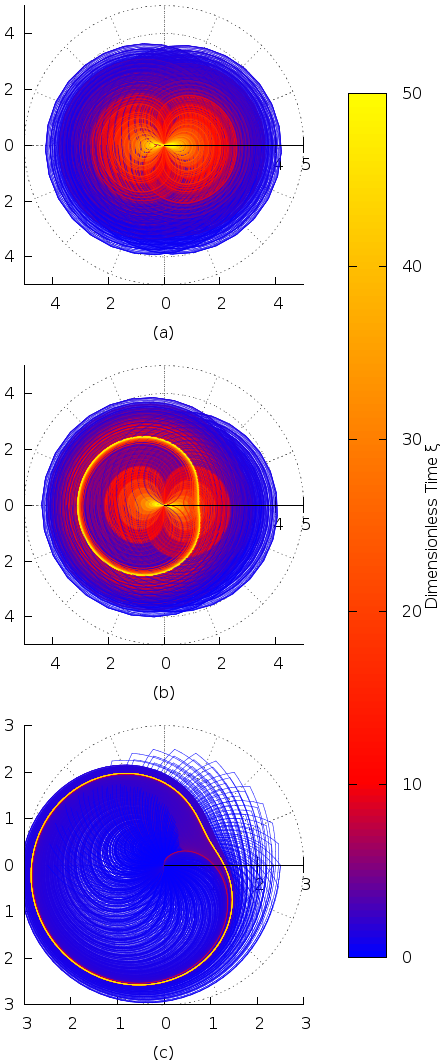}
  \centering
  \caption{(Color online) Phase portrait of \eqref{eq:josephson} in
    polar coordinate, the angle represents phase difference and the radius
    represents the time derivative of the phase difference.
    (a) $\eta=0.1,\beta=0.027$, only fixed points exist (b)
    $\eta=0.1,\beta=0.037$, bistable state: fixed points and limit
    cycle coexist (c) $\eta=0.1,\beta=1.1$, only limit cycle exist.}\label{fig:josephson}
\end{figure}
Depending on the effective dissipation coefficient $\eta$ and the
effective torque $\beta$, we can observe three kinds of bifurcation:
homoclinic, infinite-period and saddle-node bifurcations.  The
saddle-node and the homoclinic bifurcations matter here, if $\beta>1$
then there is no fixed point in the phase plane, all trajectories are
attracted to a unique, stable limit
cycle (cf. Fig.~\ref{fig:josephson}(c))\cite{Strogatz200101}; limit
cycle and stable spiral point could also coexist (bistable state
cf.~\ref{fig:josephson}(b)) if $\beta<1$ and $\eta$ is sufficiently
small, but there is no general analytical formula for the homoclinic
bifurcation curve.  A limit cycle corresponds to the unsynchronized
rotation; while when $\beta<1$, it is possible that two rotator
phase-locked (cf. Fig.~\ref{fig:josephson}(a)).

The existence of a saddle-node bifurcation point in the rotator chain
with boundary rotators will be proved in the next paragraph and
further corroborated by numerical simulation in Sec.
\ref{sec:profiles}.  

The phase difference equation of \eqref{eq:evo_simplified} is
\begin{align}
  \label{eq:full_diff_equation}
  \notag \frac{\mathrm d\delta_k}{\mathrm ds}&=\Delta_k,\\
  \notag \frac{\mathrm d\Delta_k}{\mathrm
    ds}&=-2\sin\delta_k+\sin\delta_{k-1}+\sin\delta_{k+1},&&k\neq
  1,N-1\\
  \notag \frac{\mathrm d\Delta_1}{\mathrm
    ds}&=-2\sin\delta_1+\sin\delta_2+\alpha\omega_1-\tau,\\
  \frac{\mathrm d\Delta_{N-1}}{\mathrm
    ds}&=-2\sin\delta_{N-1}+\sin\delta_{N-2}-\alpha\omega_N,
\end{align}
and the phase difference equation for two boundary rotators ($\delta_{1,N}=\phi_N-\phi_1$) is
\begin{align}
  \label{eq:diff_boundary_exclusive}
  \notag\frac{\mathrm d\delta_{1,N}}{\mathrm ds}&=\Delta_{1,N},\\
  \notag\frac{\mathrm d\Delta_{1,N}}{\mathrm ds}&=\frac{\mathrm
    d\omega_N}{\mathrm ds}-\frac{\mathrm d\omega_1}{\mathrm ds},\\
  &=-\sin\delta_{N-1}+\sin\delta_1-\alpha\Delta_{1,N}-\tau.
\end{align}
The condition $\frac{\mathrm d\Delta_k}{\mathrm ds}=0$ can be
rewritten in the form of a forward and a backward recurrent equation
\begin{align}
  \label{eq:recur_boundary}
  \notag\sin\delta_{k}&=2\sin\delta_{k-1}-\sin\delta_{k-2}&& k\geq3\\
  \sin\delta_{N-k}&=2\sin\delta_{N-k+1}-\sin\delta_{N-k+2}&&k\geq3
\end{align}
Solving $\frac{\mathrm d\Delta_{1,N}}{\mathrm ds}=0$ for
$\sin\delta_1$ and substituting the result into $\frac{\mathrm
  d\Delta_{N-2}}{\mathrm ds}=0$ gives the initial value for the backward
recurrence equation
\begin{align}
  \label{eq:initial_backward}
  \notag\sin\delta_{N-1}&=-\sin\delta_1-\tau,\\
  \sin\delta_{N-2}&=-2\sin\delta_1+\alpha\omega_N-2\tau.
\end{align}
The initial value of the forward recurrence equation is straightforward
from $\frac{\mathrm d\Delta_1}{\mathrm ds}=0$
\begin{align}
  \label{eq:initial_forward}
  \notag \sin\delta_1&=\sin\delta_1,\\
  \sin\delta_2&=2\sin\delta_1-\alpha\omega_1+\tau.
\end{align}
Then the general formulae for forward and backward recurrence equation
can be obtained
\begin{align}
  \label{eq:general_solution_recurrence}
  \notag
  \sin\delta_k&=k\sin\delta_1+(k-1)(\tau-\alpha\omega_1),\\
  \sin\delta_{N-k}&=-k\sin\delta_1-k\tau+\alpha(k-1)\omega_N,
\end{align}
and the necessary condition for fixed points then
follows($\Delta_{1,N}=\omega_N-\omega_1=0$)
\begin{equation}
  \label{eq:fixed_point_necessary}
  \sin\delta_k+\sin\delta_{N-k}=-\tau.
\end{equation}
The only solution that satisfies
\eqref{eq:fixed_point_necessary} and
$\frac{\mathrm d\delta_k}{\mathrm ds}=0$ simultaneously is
\begin{equation}
  \label{eq:solution_phase_difference}
  \sin\delta_k=-\frac{\tau}{2}.
\end{equation}
According to \eqref{eq:solution_phase_difference}, it is evident that $\tau=2$ is a saddle-node bifurcation
point, if $\tau>2$, then there could not be any
fixed point, and the entire system could not be synchronized.
Numerical simulation confirmed that $\tau>2$ is a necessary
condition for nonlinear variance profile to form.

The argument of last subsection still holds, the only differences are
$\chi=\pm\sqrt{1-\frac{\tau^2}{4}}$ and the elements in
Jacobian regarding $\Delta_{1,N}$ that render the stable fixed point a
spiral.

The above result shows under the condition $\tau<2$ the rotator chain
with dissipative boundary could have the possibility of synchronized
collective rotation.

When $\tau>2$, it can be proved that there exist trajectories of
\eqref{eq:full_diff_equation} that are confined in the energy shell
$E_r$.  $E_r$ is determined by $\frac{\mathrm dE}{\mathrm
  ds}=\omega_1(\tau-\omega_1)-\omega_N^2<0$, if $\omega_1>\tau$, then
the inequality holds under any condition, then
$E_r=N(\tau^2/2+2)$.  The determination of the types of
trajectories calls for further studies on Poincar\'e-Bendixson type
theorem in N-dimensional space.

With globally synchronous rotation, nonlinear profile of angular momenta is
impossible, since the momentum variance of each rotator is
identically $0$, which indicate $\tau>2$ is a necessary condition for
the nonlinear profile to form. 
\section{NUMERICAL SIMULATION}
\label{sec:computation}
The system of equations~\eqref{eq:evo_simplified} had been integrated
numerically by Velocity-Verlet and Gear's Predictor-Corrector
algorithm\cite{Haile199703} for a chain of 1024 rotators with the time
step size $\Delta s=0.01$, while there is some time-step bias, both
methods produced the same qualitative results with respect to the
choice of the time step.

The variance of momentum and the local energy flux are common
observables to investigate the relationship between the temperature
gradient and the energy flux in the low-dimensional system.  They are
calculated in our study to reveal the deterministic dynamics of
rotator chain.

The variance of momentum is computed by the standard definition
\begin{align}
  \label{eq:temperature}
  \mathrm{var}\{\omega_k\}&=\langle(\omega_k-\langle\omega_k\rangle)^2\rangle=\langle\omega_k^2\rangle-\langle\omega_k\rangle^2\\
  \notag&=\lim_{s\to\infty}\frac{1}{s}\int_0^s\omega_k^2(\zeta)\mathrm
  d\zeta-\left[\lim_{s\to\infty}\frac{1}{s}\int_0^s\omega_k(\zeta)\mathrm d\zeta\right]^2.
\end{align}
The expression of energy flux could be derived by substituting the equation of
motion into the derivative of the local energy and comparing the
result to the continuity equation for local energy\cite{Lepri20031,Dhar2008}
\begin{equation}
  \label{eq:flux}
  j_k=\lim_{s\to\infty}\frac{1}{2s}\int_0^s[\omega_k(\zeta)+\omega_{k+1}(\zeta)]F[\phi_{k+1}(\zeta)-\phi_{k}(\zeta)]\mathrm d\zeta.
\end{equation}

We define the norm of a solution vector to explore the periodicity of
the solution. The norm of a solution vector
$\boldsymbol{V}=\binom{\boldsymbol{\phi}}{\boldsymbol{\omega}}$ is
defined to be
\begin{equation}
  \label{eq:norm}
  |\boldsymbol{V}|=\sqrt{\sum_k(\phi_k^2+\omega_k^2)}.
\end{equation}
Note that the norm has the property $|\boldsymbol{V}|=0$ if and only
if $\boldsymbol{V}=\boldsymbol{0}$, this reduces the description of
a closed phase trajectory form a high dimensional curve to one
dimensional.  Since if $|\boldsymbol{V}(s)-\boldsymbol{V}(0)|$
returns to $0$ periodically, the trjectory $\boldsymbol{V}(s)$ is a
closed orbit.

The fixed boundary conditions were also checked numerically, this is
equivalent to adding extra torques $-\sin\phi_k(k=1,N)$ on each
end. In particular, a rotator chain with both kinds of boundary
condition had the single-peaked variance profile of momenta in the
certain parameter region.

All observables were analyzed for the time steps of $10^8-10^{10}$,
after the system had relaxed to steady state.


\subsection{An Overview of the Dynamical States}
\label{sec:profiles}
\begin{figure}[hbt!]
\includegraphics[width=.5\textwidth]{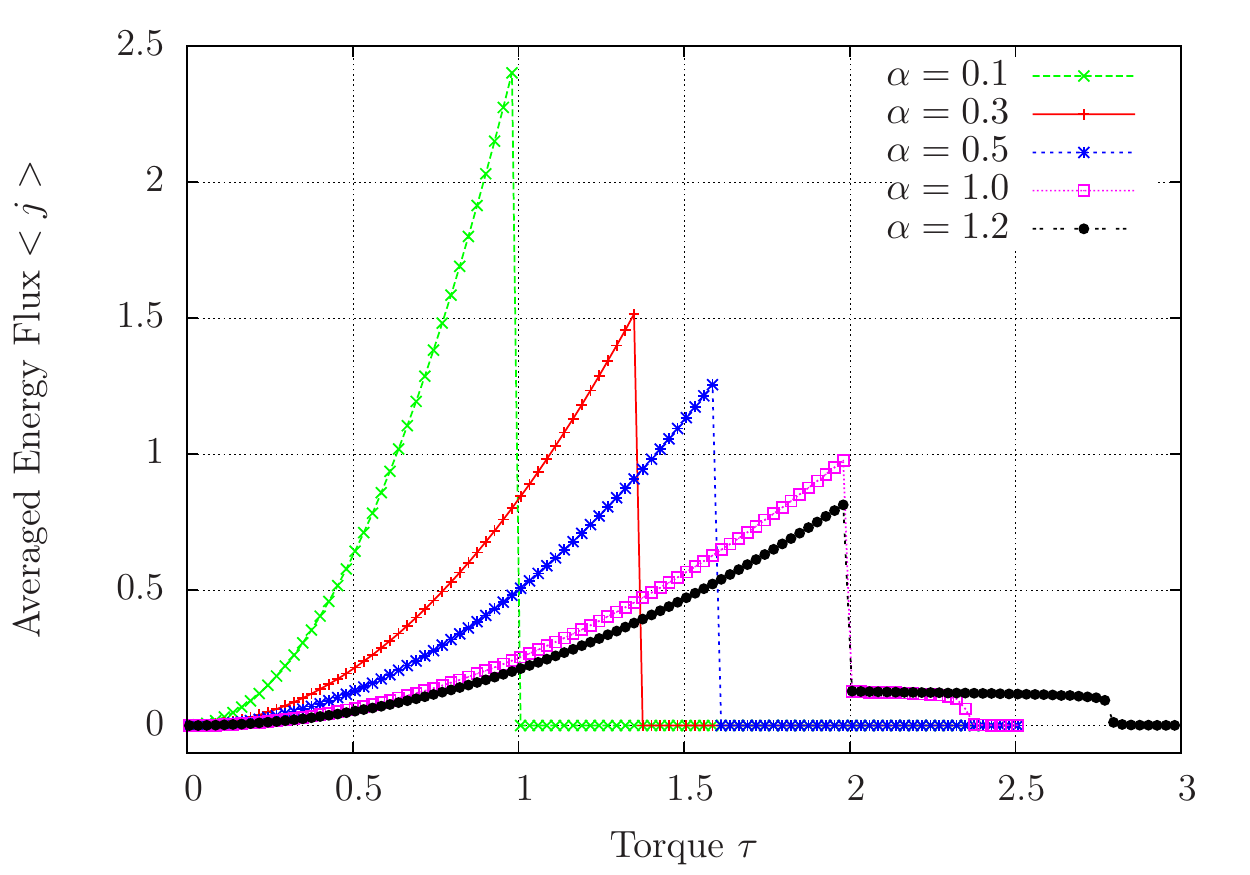}
  \centering
  \caption{(Color online) Averaged energy flux of rotator chain.}\label{fig:flux}
\end{figure}

The energy flux is plotted for revealing the overall dynamical
states of the rotator chain, see FIG.~\ref{fig:flux}.

As the driving torque increased, the torque-flux curve exhibited three
qualitatively different segments for all $\alpha>1$.  When $\tau<2$,
the local energy flux increased monotonically with driving torque as
$\langle j\rangle=\tau^2/4\alpha$, while identically plunged when
$\tau>2$, and virtually vanished if the mechanical torque is larger
than $2.636\alpha-0.437$ (by linear fit).  For the cases $\alpha<1$,
the local energy flux dropped when $\tau<2$.  We have proved rotator
chain without dissipative boundary has a saddle-node bifurcation point
$\tau=1$ (cf.  \ref{sec:interior}). As the effective dissipation
coefficient $\alpha$ decreased, $\frac{\mathrm d\Delta_1}{\mathrm ds}$
in transient state could be too large for the globally synchronous
rotation (cf. FIG.~\ref{fig:josephson}(b)), which rendered the rotator
chain in \emph{split synchronous rotation state}.

The monotonic relation between the averaged energy flux and the
driving torque is straightforward from the result derived in
Sec.\ref{sec:boundary}.  Assume that the system had relaxed to the
spiral, then $\Delta_{1,N}=\omega_N-\omega_1=0$ and
$\sin\delta_k=\tau/2$.  For the total energy $E$ of the system
$\frac{\mathrm dE}{\mathrm ds}=0$, the work done by driving torque
must be balanced by the dissipation at both end
$\tau\omega=2\alpha\omega^2$, combined with
\eqref{eq:full_diff_equation} we conclude that at spiral all rotators
have same angular momenta $\omega=\tau/2\alpha$.  Substituting this
result and \eqref{eq:solution_phase_difference} into \eqref{eq:flux}
leads to the relation $\langle j\rangle=\tau^2/4\alpha$. The proof
provides a strong evidence that the monotonically increasing segment
represent the globally synchronous rotation state of rotator chain.

\begin{figure}[hbt!]
\includegraphics[width=.5\textwidth]{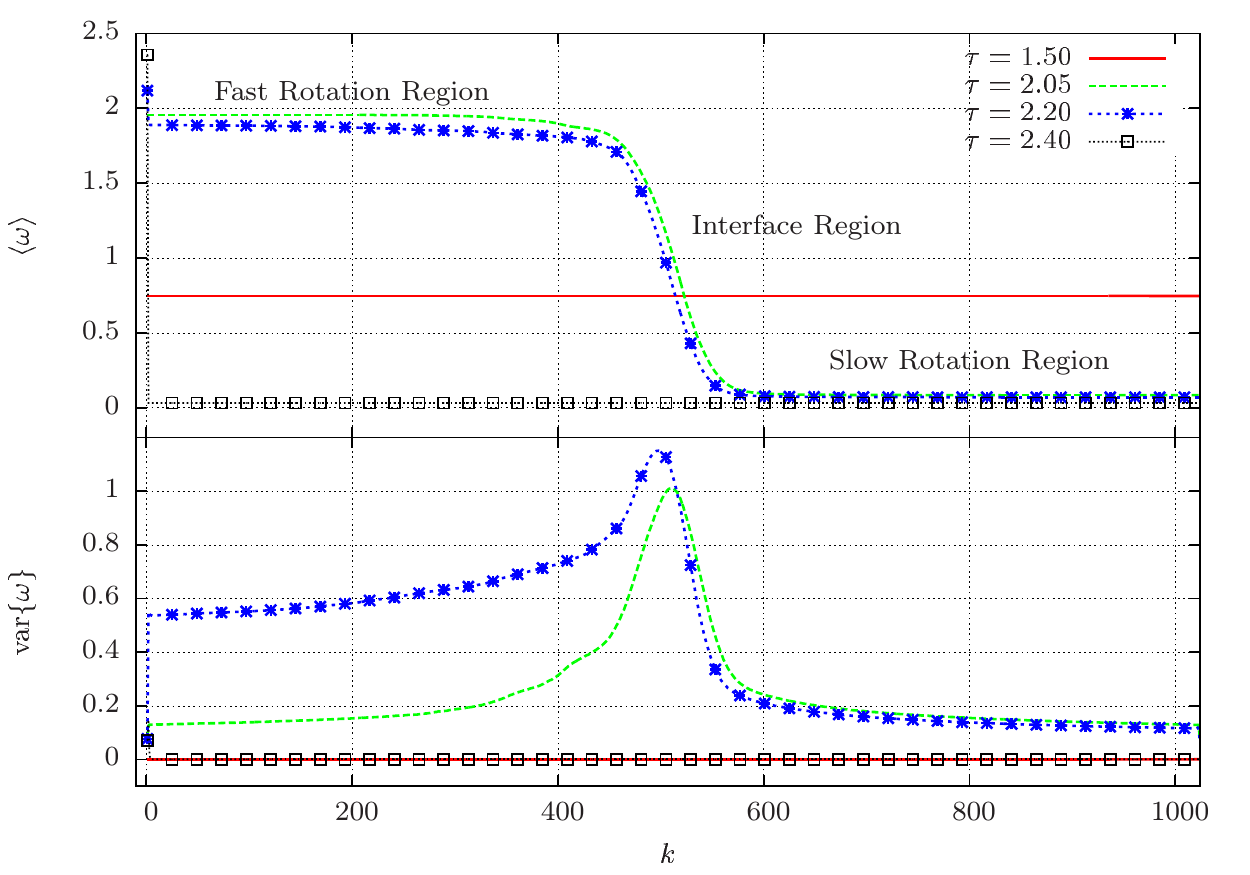}
  \centering
  \caption{(Color online)Averaged momenta and variance of momenta for chains of $\alpha=1.0$.}\label{fig:avg_variance}
\end{figure}

The averaged momenta and the variance of momenta are plotted in
FIG.~\ref{fig:avg_variance} to analyze each segment of the curve for
$\alpha=1.0$ in FIG.~\ref{fig:flux}.  As expected, the curves in
FIG.~\ref{fig:avg_variance} that represent the three segments are
essentially different. The three kinds of curves in
FIG.~\ref{fig:avg_variance} correspond to \emph{globally synchronous
  rotation}, \emph{clustered synchronous rotation} and \emph{split
  synchronous rotation} respectively.

The entire chain was synchronized when $\tau=1.5, \alpha=1.0$, as
predicted in \ref{sec:boundary} (Noting that the momentum of every
rotator are exactly $\tau/2\alpha$). As the effective torque
further increased, an interface area where averaged momenta exhibited
significant drop emerged, and the entire chain was divided into three
regions: 1. \emph{Fast rotation region} 2. \emph{Interface region} and
3. \emph{Slow rotation region}, the onset of the interface region shifted
left slightly as driving torque increased in this state. If the
driving torque is sufficiently large, the driving rotator will rotated
as if the remaining rotators on the chain do not exist (note that
$\omega\approx\tau$), this fact corresponded to the transform from
librations to rotations in the single pendulum.

The momenta variance remained $0$ when $\tau<2$ (i.e. the rotator
chain was synchronized), which confirm the results derived in
\ref{sec:dyna_state}.  When $\tau$ was sufficiently large, only
several rotators near the driving rotator had nonzero momenta
variance.  This fact brought the name \emph{split synchronous
  rotation state}, despite the first several rotators, the remaining
rotators had the same averaged momenta as well as zero variance of
momenta, the entire chain thus consists of only one fast moving
rotator while the others relaxed to slow synchronous rotation.

\begin{figure}[hbt!]
\includegraphics[width=.5\textwidth]{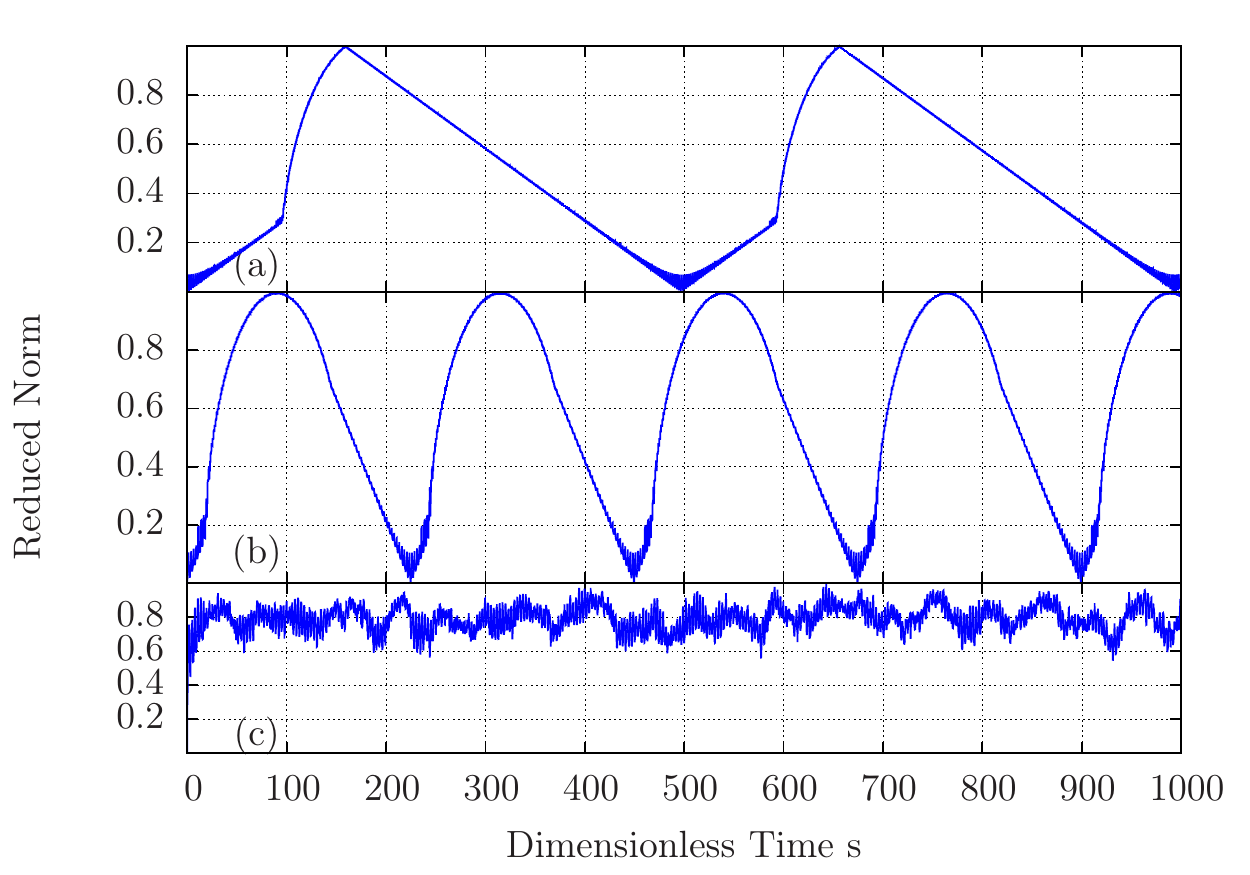}
  \centering
  \caption{(Color online)
    $|\boldsymbol{V}(s)-\boldsymbol{V}(s_0)|/\mathrm{max}\{|\boldsymbol{V}(s)-\boldsymbol{V}(s_0)|\}$
    for (a) $\tau=1.7,\alpha=0.5$, (b) $\tau=2.5,\alpha=1.0$ and (c)
    $\tau=2.2, \alpha=1.0$.)}\label{fig:closed_orbit}
\end{figure}

The phase trajectory of split synchronous rotation state is a closed
orbit while the clustered synchronous rotation state is not.  This can be shown
by calculating the reduced norm
$|\boldsymbol{V}(s)-\boldsymbol{V}(s_0)|/\mathrm{max}\{|\boldsymbol{V}(s)-\boldsymbol{V}(s_0)|\}$,
as shown in FIG.~\ref{fig:closed_orbit}.  The first two figures
represent split synchronous states, and the last one represents
clustered synchronous rotation state.

It is interesting to note that the single-peaked distribution of
variance existed when the rotator chain was in the \emph{clustered
  synchronous rotation} state and the peaks located in the interface
region.  In previous study, this form of variance profile was thought
as a consequence of the interaction between the thermal and mechanical
driving\cite{PhysRevE.84.061108}.  However in our model, thermal bath
was clearly absent, hence the cause of single-peaked variance profile
calls for further investigation by exploring the deterministic
dynamical properties of the interface rotators.



\subsection{Rotation States of Interface Rotators}
\label{sec:states}
\begin{figure}[hbt!]
\includegraphics[width=.5\textwidth]{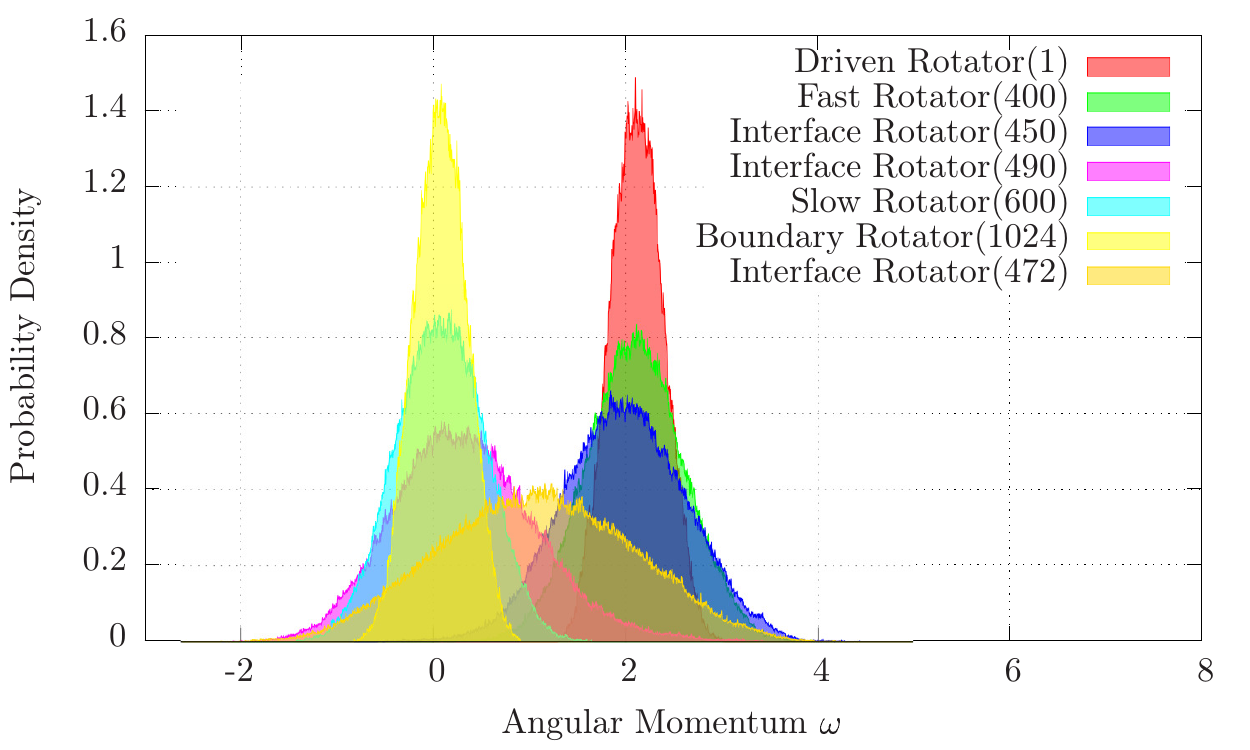}
  \centering
  \caption{(Color online) Distribution of angular momenta for
    $\tau=2.2,\alpha=1.0$.)}\label{fig:stat_mom}
\end{figure}
In order to trace the origin of the large variance on the chain it is
reasonable to analyze the momentum distribution of the typical rotators.
As shown in FIG.~\ref{fig:stat_mom}, the typical rotators in the fast
rotation region(e.g 1 and 400) had angular momenta distributions that
was symmetrical about the maximum value of angular momentum whereas
those in the slow rotation region (e.g 600 and 1024) had angular momenta
distributions around zero.  It is interesting to note that the
rotators in the interface region (450, 472 and 490) had a momenta
distribution extended from the slow rotation region to the fast rotation
region, which provided the evidence that the rotators locate in the
interface region constantly switched between the slow rotation state
and the fast rotation state, which result in large variance.

\begin{figure}[hbt!]
\includegraphics[width=.5\textwidth]{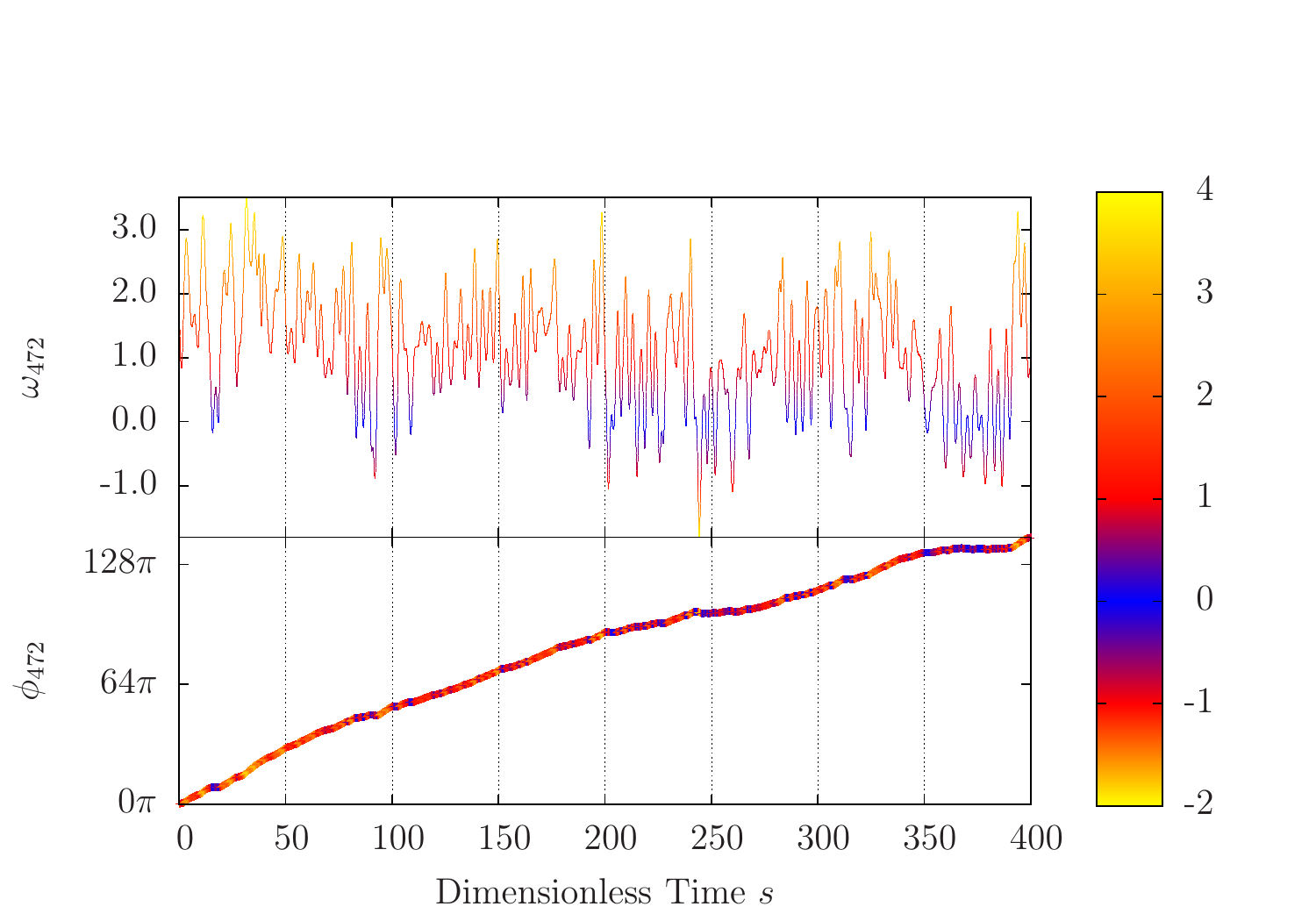}
  \centering
  \caption{(Color online) Evolution of angle and angular momentum (rotator 472) for
    $\alpha=1.0, \tau=2.2$.}\label{fig:evo_momenta}
\end{figure}

For a clearer view, the evolution of angular momentum for an interface
rotator is plotted in FIG.~\ref{fig:evo_momenta}.  The oscillation
between fast rotation and slow rotation can then be easily spotted.

\begin{figure}[hbt!]
\includegraphics[width=.5\textwidth]{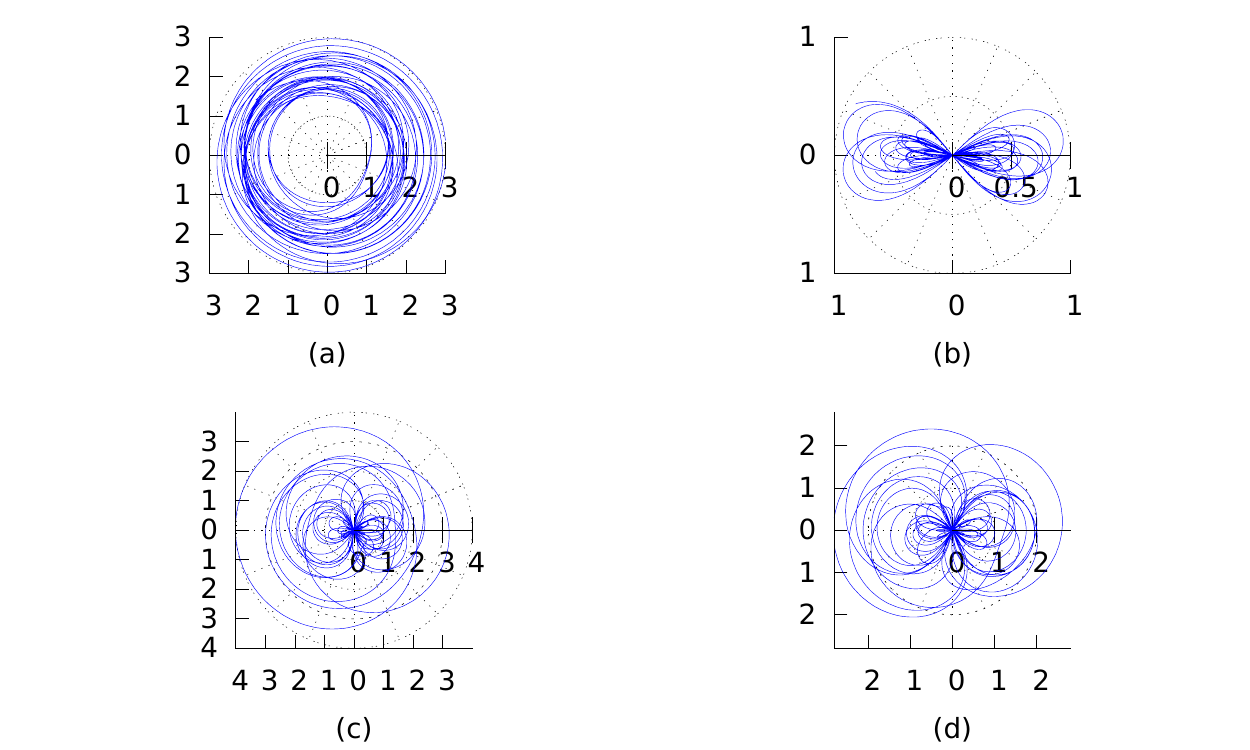}
  \centering
  \caption{(Color online) Phase portrait for phase difference between (a)
    two boundary rotators (b) two adjacent rotators in fast rotation
    region (c) interface rotator 472 and rotator 471 (d) interface
    rotator 472 and 473 ($\alpha=1.0,\tau=2.2$).}\label{fig:polar_diff}
\end{figure}

It has been proved in \ref{sec:boundary} that the boundary rotators
can not be synchronized when $\tau>2$, as illustrated in
FIG.~\ref{fig:polar_diff}(a).  Nevertheless, the adjacent interior
rotators had different synchronization states.  For the adjacent
rotator pairs in either the fast rotation region and the slow rotation
region, the phase difference and its derivative oscillate around
$(0,0)$, which corresponded to the librations in single pendulum (See
FIG.~\ref{fig:polar_diff}(b)).  The case of rotator pairs in the
interface is more complicated, the phase difference between both left
and right rotators constantly shifted between librations to rotations
and vice versa (See FIG.~\ref{fig:polar_diff}(c) and (d)), which
accounted for the oscillations between fast rotation and slow rotation
in the interface rotators. Since the rotators in both fast rotation
region and the slow rotation region are separately synchronized, and the
interface rotators temporarily synchronized to both side, hence the
name \emph{clustered synchronous rotation} is designated.
\section{CONCLUSION}
\label{sec:conclusions}
The investigations presented above demonstrates that the dynamics of
rotator chain with mechanical driving and dissipative boundary depends on
the effective driving torque $\tau$ and the effective dissipation coefficient $\alpha$.

Three dynamical states are identified for $\alpha>1$.  \emph{globally
  synchronous rotation} is proved to exist when $\tau<2$, i.e. the
driving torque is small while the coupling is strong. \emph{split
  synchronous rotation} state emerges when $\tau$ is sufficiently
large so that no other rotators on the chain could be synchronize with
the driving rotator. \emph{clustered synchronous rotations} result
from interior rotators' partial synchronization to each boundary.

The existence of the single-peaked variance profile of momenta is
confirmed in the absence of heat bath, and the large variance in the
interface region is the consequence of clustered synchronous rotation
states, which emerge when $N>2$.  Temperature is commonly defined as
variance of momenta in the research of heat conduction problem in low
dimensional system.  But as our study shows, the variance profile of
momenta could be the consequence of deterministic dynamics
exclusively, and has no implication with heat (i.e. stochastic
dynamics).  Moreover, a rotator chain with mechanical driving has
three kinds of dynamical states rather than mere oscillation in the
case of oscillator-based systems.  Thermal driving could induce
transitions between these states, thus whether the operational
definition of temperature used in oscillator-based system is
appropriate in rotator chain still calls for further investigation.

The deterministic dynamics is essential for understanding the heat
conduction properties of the rotator chain when heat baths are added
to the system. Heat baths turn the equations of motion of the system
from odinary differential equations to stochastic differential
equations. Recent developments on stochastic process reveals that
there are correspondences between the deterministic dynamics of a
system and its stochastic counterpart when a specific stochastic
interpretation other than It\^o and Stratonovich is
used\cite{1742-5468-2012-07-P07010,PhysRevE.87.062109,ma2012potential}.
As a result, it is reasonable to investigate the transport properties
of rotator chain by calculating the steady state distribution of the
model based on the dynamics of its deterministic counterpart.

\section*{ACKNOWLEDGMENTS}
\label{sec:acknowledgements}
This work has been financially supported by grants from the National
Natural Science Foundation of China (11075016) and the Foundation for
Doctoral Training from MOE.


\bibliography{ref_data}
\end{document}